\documentclass[sigconf]{acmart}

\usepackage{enumitem}
\usepackage{multirow}
\usepackage{color}
\definecolor{olive}{rgb}{0.1,0.8,0.3}
\definecolor{mauve}{rgb}{0.48,0,0.72}

\newcommand{\np}[1]{{\color{black}\textbf{}#1}\normalfont}

\AtBeginDocument{%
  }

\copyrightyear{2024}
\acmYear{2024}
\setcopyright{rightsretained}
\acmConference[CHI '24]{Proceedings of the CHI Conference on Human Factors in Computing Systems}{May 11--16, 2024}{Honolulu, HI, USA}
\acmBooktitle{Proceedings of the CHI Conference on Human Factors in Computing Systems (CHI '24), May 11--16, 2024, Honolulu, HI, USA}
\acmDOI{10.1145/3613904.3642268}
\acmISBN{979-8-4007-0330-0/24/05}

\acmConference[CHI'24]{Conference on Human Factors in Computing Systems}{May 11--16,
  2024}{O'ahu, Hawai'i, USA}
\acmBooktitle{CHI'24: ACM CHI Conference on Human Factors in Computing Systems,
 May 11--16, 2024, O'ahu, Hawai'i, USA} 

\begin{document}

\title{Stretch your reach: Studying Self-Avatar and Controller Misalignment in Virtual Reality Interaction}

\author{Jose Luis Ponton}
\email{jose.luis.ponton@upc.edu}
\orcid{0000-0001-6576-4528}
\affiliation{%
  \institution{Universitat Politècnica de Catalunya}
  \city{Barcelona}
  \country{Spain}
}
\author{Reza Keshavarz}
\email{reza.keshavarz@studio.unibo.it}
\orcid{0009-0002-5951-9459}
\affiliation{%
  \institution{Università di Bologna}
  \city{Bologna}
  \country{Italy}
}
\author{Alejandro Beacco}
\email{alejandro.beacco@upc.edu}
\orcid{0000-0001-8192-1431}
\affiliation{%
  \institution{Universitat Politècnica de Catalunya}
  \city{Barcelona}
  \country{Spain}
}
\author{Nuria Pelechano}
\email{npelechano@cs.upc.edu}
\orcid{0000-0002-1437-245X}
\affiliation{%
  \institution{Universitat Politècnica de Catalunya}
  \city{Barcelona}
  \country{Spain}
}

\renewcommand{\shortauthors}{Ponton, et al.}

\begin{abstract}
Immersive Virtual Reality typically requires a head-mounted display (HMD) to visualize the environment and hand-held controllers to interact with the virtual objects. Recently, many applications display full-body avatars to represent the user and animate the arms to follow the controllers. Embodiment is higher when the self-avatar movements align correctly with the user. However, having a full-body self-avatar following the user's movements can be challenging due to the disparities between the virtual body and the user's body. This can lead to misalignments in the hand position that can be noticeable when interacting with virtual objects.
In this work, we propose five different interaction modes to allow the user to interact with virtual objects despite the self-avatar and controller misalignment and study their influence on embodiment, proprioception, preference, and task performance. We modify aspects such as whether the virtual controllers are rendered, whether controllers are rendered in their real physical location or attached to the user's hand, and whether stretching the avatar arms to always reach the real controllers. We evaluate the interaction modes both quantitatively (performance metrics) and qualitatively (embodiment, proprioception, and user preference questionnaires). Our results show that the stretching arms solution, which provides body continuity and guarantees that the virtual hands or controllers are in the correct location, offers the best results in embodiment, user preference, proprioception, and performance. Also, rendering the controller does not have an effect on either embodiment or user preference. 

  
\end{abstract}



\begin{teaserfigure}
  \includegraphics[width=\textwidth]{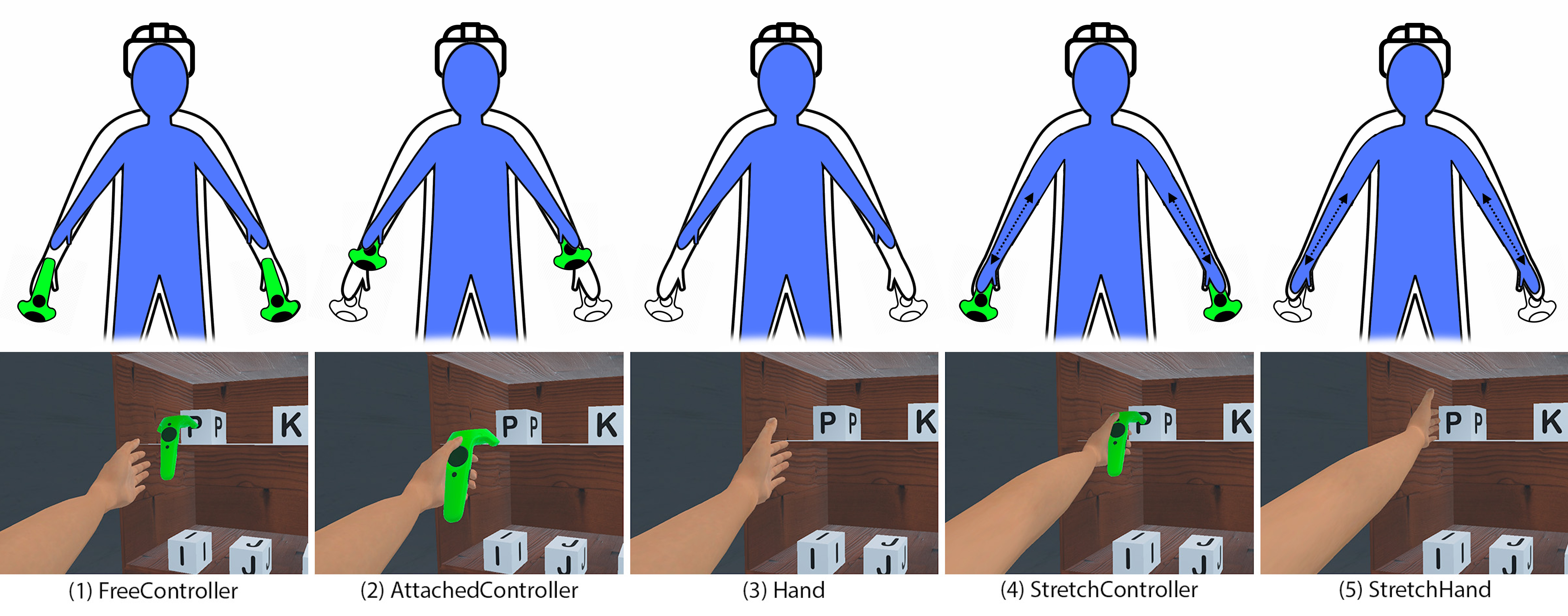}
  \caption{
  Five interaction modes are proposed to manipulate virtual objects when the user has a full-body avatar animated with Inverse Kinematics. (1) \emph{FreeController}, the controller is rendered and may appear to fly when the avatar's arm is shorter than the user's arm; (2) \emph{AttachedController}, the controller is rendered always attached to the user's virtual hand even when the avatar's arm is shorter than the user's arm; (3) \emph{Hand}, the controller is not rendered and thus the virtual hand reach depends on the avatar's arm length; (4) \emph{StretchController}, the controller is rendered and the arms stretch to always match the position of the virtual and real hand, thus always holding the virtual controller; similarly (5) \emph{StretchHand} has the same behavior but without rendering the controller. For a summary of all methods see Table \ref{tab:modes}. 
  }
  \Description{The figure illustrates five interaction modes for manipulating virtual objects with a full-body avatar using inverse kinematics. Each mode is represented by a simplified human and avatar figures with a green virtual controller in their right hand. The modes are: (1) \emph{FreeController}, the controller is rendered and may appear to fly when the avatar's arm is shorter than the user's arm; (2) \emph{AttachedController}, the controller is rendered always attached to the user's virtual hand even when the avatar's arm is shorter than the user's arm; (3) \emph{Hand}, the controller is not rendered and thus the virtual hand reach depends on the avatar's arm length; (4) \emph{StretchController}, the controller is rendered and the arms stretch to always match the position of the virtual and real hand, thus always holding the virtual controller; similarly (5) \emph{StretchHand} has the same behavior but without rendering the controller.}
  \label{fig:modes}
\end{teaserfigure}


\sloppy
\maketitle

\section{Introduction}
\label{sec:intro}

Immersive Virtual Reality is a technology with the potential to revolutionize fields like education, healthcare, and entertainment, offering immersive, impactful experiences. Having users embodied with a virtual avatar can improve spatial perception, among others, and has already been used for applications in learning, rehabilitation, and mental health. 
Immersive VR can make the users feel as if they are actually present in a virtual world and behave as if they were in the real world.. 
A virtual body representation that follows the user's moves with visuomotor synchrony can provide the illusion of body ownership, leading to embodiment \citep{kilteni:2012}, which can enhance presence \citep{slater:2023}. Ideally, the virtual avatar should perfectly mimic the user to facilitate natural interactions since inaccuracies in body pose can reduce embodiment and performance \cite{Yun:2023}


Given the challenges of having accurate avatar representations when wearing a Head Mounted Display (HMD), often, the user is represented with a floating upper torso and hands.
In these cases, the upper body is located under the HMD position, and the virtual hands are rendered at the position of the handheld controllers, and provide a way to interact with the virtual world. However, when rendering a full virtual body, it is often difficult to accurately co-locate the virtual and real hand positions due to limitations of the virtual body, such as body size or proportions \cite{Ponton:2022}, skeleton simplification, or inaccurate offsets with respect to trackers \cite{Ponton:2023}. Respecting the virtual body dimensions may lead to misalignments between the virtual hand/controller and the real hand/controller when reaching far-away objects.


Body discrepancies may not be perceptually noticeable in applications that do not require careful manipulation of objects, for example, those where the user has to move freely \cite{Ponton:2023}, walk \citep{llobera:2021}, gesticulate while talking \citep{slater:2018}, or even hold a controller and eventually press a button \citep{neyret:2020}. 
However, when direct manipulation is needed while rendering a full avatar, it may be necessary to redirect the hand position to respect the limits imposed by the virtual body. 
Virtual hand redirection, when rendering only hands or hands with part of the arm, has been studied to improve ergonomics \cite{Murillo:2017}, to induce the perception of weight in virtual objects \cite{Rietzler:2018}, or to induce a proprioceptive drift and a change in body schema \cite{kilteni:2012b}. However, when a full virtual body is being rendered, altering the position of the virtual body with respect to the user could affect embodiment. 
The work by \citet{Tao:2023} improved the realism of the interactions between the self-avatar and the environment by adding physics to the virtual body. This led to discrepancies between the virtual and the real body pose, which could increase or reduce embodiment depending on the intensity. This work set the basis to continue researching to what extent we can introduce discrepancies to improve interaction in the virtual world without reducing embodiment. In this paper, we focus on interactions between the hand and the virtual objects.



Current solutions that can be found in most applications either render only floating hands always on the controller position \cite{botvinick:1998, Burns:2005, argelaguet:2016}, or not render the controllers when having a full avatar \cite{dewez:2021, goncalves:2022, kober:2022}, to avoid the misalignments with the virtual hands caused by the limitations imposed by having a full-body avatar.




Not rendering the controllers may be enough for VR applications with little manipulation of objects, such as applications to imitate an avatar to learn to dance or practice yoga, or social applications providing non-verbal communication.
However, if the virtual controllers are required for interaction purposes (e.g., used as a tool to interact closely with objects), or, similarly, if a virtual tool must be rendered, then certain trade-offs arise. For example, suppose the controllers are rendered in their real-world position, and the user has a full-body avatar. In that case, users may be trying to interact with a virtual object that is far away (users need to stretch the arm) and observe the virtual controller moving away from the avatar's hand \citep{Ponton:2022} (as if it was flying because the virtual arm cannot reach such position). This can be perceptually noticeable for users as they may notice an inconsistency between haptic feedback (feeling the controller's touch in the hand) and visual feedback (observing the controller flying away from the virtual hand). 

One solution could be to render the controller always attached to the avatar's hand to achieve multisensory feedback (visual and haptic). Then, by the induced embodiment, we may convince the users that their body has changed dimensions to accept the modified body as their own \citep{kilteni:2012b}. 
It could then be possible to have believable interaction with the virtual objects in the environment despite the misalignments between the virtual body and the user. In some studies, it has been observed that this may lead to a proprioceptive drift which makes the user move toward the avatar pose to fill the gap \citep{Gonzalez:2020}. However, when trying to grab out-of-reach objects, it is impossible to close the gap and may lead to a proprioceptive conflict. Another alternative could be to stretch the avatar's arm so that the virtual hand can always hold the controller \citep{feuchtner:2017}, although the unnatural skinning could negatively affect appearance, which is also important for embodiment \cite{argelaguet:2016, fribourg:2020}.


Our goal is to study interaction metaphors that can be used as solutions to overcome current limitations when animating a full virtual avatar to represent the user while interacting with the environment.
In this work, we propose and explore five interaction metaphors consisting of different representations for the user when interacting with virtual objects. These representations are different combinations of binary factors, such as rendering the virtual controllers, attaching them to the virtual hands, and stretching the virtual arms. 
Our specific research questions are the following: 
\begin{itemize}
    \item \textbf{RQ1:} How does the rendering of virtual controllers impact user embodiment and performance in object manipulation tasks within virtual reality?
    \item \textbf{RQ2:} How does the accuracy of virtual controllers' absolute location affect user embodiment and performance in object manipulation tasks within virtual reality?
    \item \textbf{RQ3:} How does attaching the virtual controllers to the virtual hand impact user embodiment and performance in object manipulation tasks within virtual reality?
\end{itemize}
The goal of our study is to determine which of our interaction metaphors is preferred by users and can enhance embodiment and performance. All our methods have the same avatar dimensions corresponding to a uniform scaling of an avatar based on the participant's height, and, thus, is not an exact match of the participant's proportions. 
The contributions of this work are: 
\begin{itemize}
    \item A study of five distinct interaction metaphors, focusing on three key factors: \emph{Controller}, \emph{Attached}, and \emph{Stretch}. We analyze their relationship with embodiment, performance, user preference, and proprioception in virtual reality. A significant finding from our study is that stretching arms consistently yields the best results across all these metrics. Furthermore, visually rendering VR controller does not significantly impact these outcomes. 
    \item Based on the insights from the study, we formulate a set of guidelines for designing manipulation techniques for full-body self-avatars in virtual reality.
\end{itemize}


\section{Related work}

Head-mounted displays for Virtual Reality are increasingly becoming popular for performing some tasks due to the exceptional degrees of immersion they afford. Specifically, tasks involving high interactivity levels seem to improve the sense of presence, thereby enhancing the overall user experience \citep{vergari:2021}.
There is extensive literature focused on discerning methods to quantify embodiment, presence, and immersion in VR. 
Probably the most relevant of those studies is the rubber hand illusion \citep{botvinick:1998}, which examines the sense of ownership over a rubber hand synchronized with visual and tactile stimuli, leading to the illusion that the rubber hand belongs to the participant. This feeling of ownership can also be replicated in a virtual environment \citep{slater:2008}, even without tactile stimulation \citep{sanchez-vives:2010}.

Seeking to better understand and classify the mechanisms that generate these illusions, \citet{kilteni:2012} delves into the Sense of Embodiment (SoE) concept, defined as perceiving an external body as one's own. They propose three key dimensions influencing SoE: the Sense of Self-Location (SoL) pertaining to the feeling of being within the virtual body, the Sense of Agency (SoA) regarding the perception of control over the virtual body, and the Sense of Body Ownership (SoO) about the feeling that the virtual body is one's own.

Given the growing interest in the SoE, \citet{gonzalez-franco:2018} conducted a comprehensive review of embodiment questionnaires, resulting in a standardized questionnaire for embodiment experiments. This questionnaire encompasses six primary question types to assess embodiment, namely, body ownership, agency and motor control, tactile sensations, body location, external appearance, and response to external stimuli. Later, \citet{peck:2021} further refined this questionnaire, proposing four key embodiment components: appearance, response, ownership, and multi-sensory. While agency, touch, and localization are not explicitly included in this categorization, the researchers discovered that these components contribute to the four major embodiment categories as they correlate with other senses.

\subsection{Body Representation in Virtual Reality}
While numerous factors, such as the visual rendering of the virtual environment or the degree of interactivity \citep{vergari:2021}, influence the SoE, a crucial contributor is the representation of the user's body via an avatar\textemdash commonly termed a self-avatar. This representation considerably augments SoE and proves advantageous for different tasks like egocentric distance estimation, spatial reasoning, and collision avoidance \citep{ries_Analyzingeffectvirtual_2009,pan_AvatarTypeAffects_2019,pan_HowFootTracking_2019}.

A variety of studies have examined how different aspects of a full-body avatar impact SoE. For instance, \citet{goncalves:2022} analyzes the influence of varying numbers of tracking points and locomotion, whereas \citet{fribourg:2020} investigates the effect of point of view, appearance, and control on SoE and its components. However, the impact of the interaction between the full-body avatar and the virtual objects on SoE remains unexplored.

To address this gap, several libraries have been introduced to augment embodiment with full-body avatars \citep{Ponton:2022, oliva:2022}. In our study, we go one step further: we instantiate the user in a full-body self-avatar to create a convincing embodiment illusion and then examine how the representation of virtual controllers, hands, and their positioning in the virtual environment\textemdash in relation to their real counterparts\textemdash influence interaction and immersion.

\subsection{Redirected hand position}
Hand redirection has been used in immersive VR with several goals. \citet{Rietzler:2018} introduced perceivable tracking offsets between the user's hand and the virtual hand in order to simulate weight when carrying objects. \citet{Murillo:2017} developed a redirection technique to improve ergonomics by reducing the range of physical movements needed to reach virtual objects within a virtual environment. Their method takes advantage of the dominance of the human visual system over the proprioceptive system. \citet{hoyet2023stick} presented two methods to recover the position mismatch between users’ real and virtual hands after releasing contact with a virtual object. \citet{Gonzalez:2023} proposed a sensorimotor model of hand redirection to simulate several movement features such as hand trajectory or velocity. Their model could be used as a tool to evaluate hand redirection techniques without the need for user testing.

The main difference between previous work on redirected hand position is that they render either just a hand or a hand with part of the arm. Still, without a full virtual body or else the virtual body is mostly static and only the hand can perform small movements. Therefore, they do not need to worry about embodiment or body continuity when the user has complete freedom of movement. The challenge when rendering a hand in a different position while having a full avatar is that it may break the body ownership illusion, which will affect embodiment. Previous work has shown that when elongating the arm length, users can feel body ownership, although the illusion diminishes with the length of the arm \citep{kilteni:2012b}. The effect is induced by the congruence visuo-motor and visuo-tactile feedback, even if there are incongruences between proprioception and visual feedback (i.e., the hand being rendered far away from the real hand). However, in their work, the user could only sit still resting the arm on a table while doing small movements to touch the surface.

\subsection{Interaction Modes in Virtual Reality}
Interaction in VR is crucial to enhance the levels of immersion, necessitating meticulous design to bolster user performance within virtual environments. Typically, VR interaction is facilitated by monitoring the user's hand movements via a device, such as a VR controller or a vision-based hand-tracking mechanism.

Early research conducted by \citet{argelaguet:2016} explored the impact of virtual hand rendering on the SoE for a pick-and-place task. The results indicated that a higher SoA was achieved with more simplistic models, such as a sphere, as compared to a realistic hand model. However, the potential inaccuracy in hand tracking may have influenced these findings. In line with this, several subsequent studies compared multiple methods to render a virtual hand. \citet{lin:2016} found that all hand rendering types could induce an illusion of body ownership, although less effectively, with non-anthropomorphic models. \citet{ricca:2021} examined the role of hand visualization in tool-based training, finding no significant difference in performance when hands were rendered versus when they were not.

Subsequently, some researchers shifted their focus to the rendering of the virtual arm. \citet{tran:2017} reported that more basic representations (such as only hand or hand and wrist) performed faster in their tasks, yet no significant differences were found in accuracy, SoA, and SoO. \citet{kober:2022} explored the EEG activity when rendering a realistic hand and arm representation versus a skeleton-based one, finding that similar activations were achieved with the more realistic arm model.

Despite these advancements, research into hand or arm rendering has led to varied results, suggesting the type of task being performed may be a critical factor, and thus, more research in this area is needed. Given the potential influence of hand tracking accuracy on these outcomes \citep{dewez:2021}, some studies have compared hand tracking systems with 6-DoF virtual reality controllers \citep{caggianese:2019, gusai:2017, depaolis:2020, depaolis:2022}. Here, all studies found that the use of VR controllers enhanced performance, user experience, and embodiment.

Although VR controllers are commonly used for VR interaction, there are few studies on their effects on the participants. There is, therefore, a need to expand knowledge about VR interactions involving controllers. In response, \citet{gao:2023} developed a general questionnaire for evaluating interactions with objects across various VR applications. \citet{lougiakis:2020} compared three methods of rendering controllers\textemdash a sphere, a virtual controller, and a hand. They found no significant differences in SoA; however, the sphere under-performed significantly, and the controller outperformed the other conditions in the positioning task. Other studies compared the use of VR controllers for rehabilitation \citep{juan:2023}, although their focus was predominantly on the physical controller rather than the virtual representation. 

In our work, we focus on studying interaction metaphors when using controllers with a full-body avatar. Our interaction metaphors propose alternatives to solve the mismatches between the user and the avatar, due to the simplified avatar skeleton with respect to the human counterpart. 

\section{Design}

In this section, we present the different factors involved in interacting with virtual objects that we have considered to examine. By manipulating these factors, we will obtain the different interaction modes that will be part of the user study.

\subsection{Interaction modes}

When embodying the user into a full-body avatar while using hand-held controllers, there are different decisions to be made in order to achieve different embodiment configurations. Depending on the factors we consider and those decisions, some problems might arise or not.
For instance, when interacting with virtual objects, we often consider whether we should render the virtual representation of the VR controllers. While rendering the VR controllers provides accurate visual feedback, it may be detrimental if it brings up problems with virtual arm length or hand position due to size differences between the user and the avatar.
Also, if the virtual controller consistently stays rendered in the palm of the virtual hand, we are able to maintain multi-sensory feedback and body continuity. However, this may be at the expense of not correctly aligning the virtual and the real controllers when the avatar’s arms are not long enough, which can lead to confusion due to conflicting proprioceptive feedback (the user can sense that the distance to the hand is not the real one). 
An option to solve this problem when the real controller is located outside the avatar’s virtual arm reach is to stretch the arm. This provides virtual body continuity and allows the virtual controllers to be correctly aligned with the real ones.

Therefore, we have considered and examined in our study the following three critical factors:
\begin{enumerate}
    \item[(a)] $\mathit{Controller}$: whether the virtual controller is visually represented or not.
   
    \item[(b)] $\mathit{Attached}$: whether the virtual controller remains consistently positioned in the palm of the virtual hand.

    \item[(c)] $\mathit{Stretch}$: whether the avatar's arm is extended to align the virtual controller with its corresponding absolute position in the physical world.
\end{enumerate}

\begin{figure}[htb]
  \centering
  \includegraphics[width=1.0\linewidth]{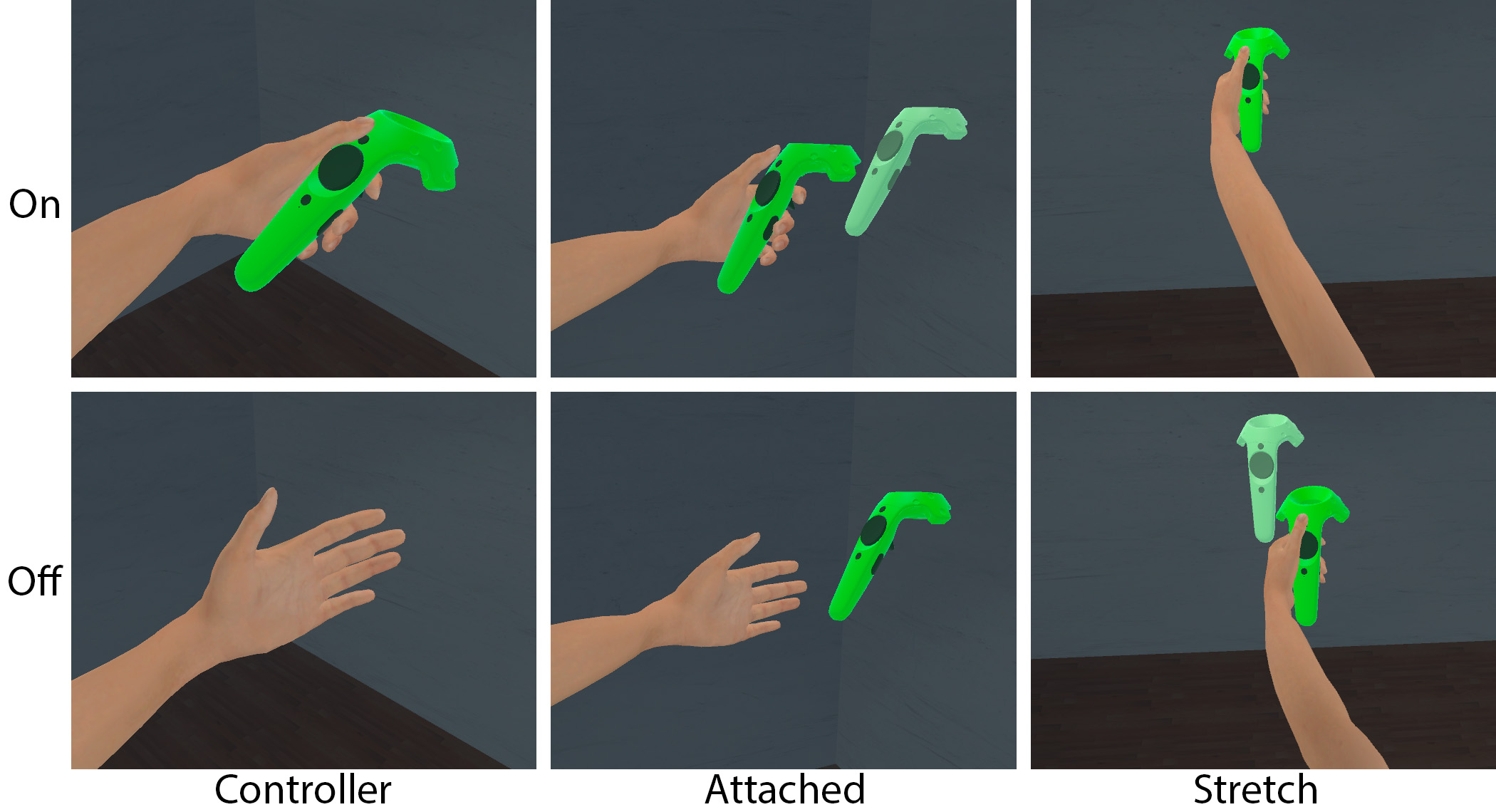}
  \caption{Illustration of the different factors in two scenarios. The top row represents the scenario in which the factor is enabled, and the bottom row corresponds to the disabled factor. From left to right, the factors $\mathit{Controller}$, $\mathit{Attached}$, and $\mathit{Stretch}$ are shown. In situations where the virtual controller is not aligned with the real one, the real controller is depicted in the figure with a lighter color (not in the simulation).}
  \Description{This figure shows six images in two rows depicting different factors in two scenarios for virtual controller manipulation. The top row indicates the 'On' scenario where the factor is enabled, while the bottom row shows the 'Off' scenario where the factor is disabled. From left to right, the factors are labeled 'Controller', 'Attached', and 'Stretch'. In each image, a human arm is shown reaching out to a virtual green controller. In the 'Controller' factor, the virtual controller is rendered in the 'On' scenario, and not rendered in the 'Off' scenario. For the 'Attached' factor, the virtual controller appears attached to the hand in the 'On' scenario, and detached in the 'Off' scenario. In the 'Stretch' factor, the arm is elongated to reach the controller in the 'On' scenario, while in the 'Off' scenario, the arm is not stretched, showing the real controller in a lighter color not aligned with the virtual one.}
  \label{fig:factors}
\end{figure}

These three binary factors (see Fig.~\ref{fig:factors}) are used to guide the design of our interaction metaphors that we have devised and implemented to study the research questions outlined in Section~\ref{sec:intro}.
The manipulation of the $\mathit{Controller}$ factor directly influences the study of \textbf{RQ1}. However, the remaining factors contribute to both \textbf{RQ2} and \textbf{RQ3}. The $\mathit{Attached}$ factor directly impacts \textbf{RQ2}; nonetheless, forcing the virtual controller always to be attached to the palm of the virtual hand may introduce misalignment issues as the avatar's arm may not perfectly match the user's arm. To address this concern, we introduced the $\mathit{Stretch}$ factor, which allows us to maintain the accurate absolute positioning of the virtual controller while also ensuring its attachment to the hand.

Considering all possible combinations of these three factors would result in eight distinct interaction modes. However, we carefully selected five interaction modes to ensure validity within the VR domain. Hence, there are three invalid combinations: firstly, not attaching the controller to the hand when the arm is stretched; secondly, when the controller is not rendered it does not matter whether the virtual controller is attached or not, consequently eliminating two more modes. 

The final five interaction modes ($\mathit{Mode}$) are summarized in Table~\ref{tab:modes} and visualized in Fig.~\ref{fig:modes}. (1) \emph{FreeController} renders the controller in its real-world position but can break body continuity and visual-tactile sensory feedback. (2) \emph{AttachedController} positions the controller in the avatar's hand, ensuring body continuity and multi-sensory cues, but may not be in its real-world position thus leading to a proprioceptive conflict. (3) \emph{Hand} mode, where the VR controller is not rendered, might not always align with the real-world hand position due to the virtual avatar's arm being shorter than the user's, thus also leading to a proprioceptive conflict. (4) \emph{StretchController}, stretches the avatar's arm to reach the real-world controller location when necessary and always renders the controller attached to the hand. Finally, (5) \emph{StretchHand}, is similar to \emph{StretchController}, but without rendering controller.

Although not initially employed to design the interaction modes, we introduce a fourth auxiliary binary factor to facilitate the study of proprioception for our interaction modes:

\begin{enumerate}
    \item[(d)] $\mathit{Location}$ determines when the virtual controller is rendered, whether its position remains aligned with its corresponding absolute position in the physical world. If the virtual controller is not rendered, $\mathit{Location}$ refers to whether the virtual hand position was aligned or not with the real hand position. 
\end{enumerate}

The inclusion of the $\mathit{Location}$ factor will serve to isolate the impact of the precise virtual controller or hand positioning, independent of other factors that primarily affect rendering aspects. 

\begin{table}[htb]
\begin{tabular}{|l|cccc|}
\hline
\multirow{3}{*}{\centering\textbf{Mode}} & \multicolumn{4}{c|}{\textbf{Factor}} \\
\cline{2-5}
& \small{$\mathit{Controller}$} & \small{$\mathit{Attached}$} & \small{$\mathit{Stretch}$} & \small{$\mathit{Location}$} \\
& (a) & (b) & (c) & (d) \\
\hline
(1) \small{\emph{FreeController}} & Yes & No & No & Yes \\
(2) \small{\emph{AttachedController}} & Yes & Yes & No & No \\
(3) \small{\emph{Hand}} & No & N/A & No & No \\
(4) \small{\emph{StretchController}} & Yes & Yes & Yes & Yes \\
(5) \small{\emph{StretchHand}} & No & N/A & Yes & Yes \\
\hline
\end{tabular}
\caption{Interaction modes used in the study. Four binary factors are used to determine the behavior of each mode: (a)~$\mathit{Controller}$, (b)~$\mathit{Attached}$, (c)~$\mathit{Stretch}$, and (d)~$\mathit{Location}$. The interaction modes can be seen in Fig.~\ref{fig:modes}.}
\label{tab:modes}
\end{table}

\subsection{Interaction Tasks}
\label{sec:tasks}
We designed three distinct interaction tasks encompassing various upper-body actions to evaluate task performance and user immersion in the self-avatar. These tasks aimed to provide users with various upper-body interactions manipulating virtual objects and enable them to assess the level of immersion achieved.

\paragraph{\textbf{Cube Task}}
Participants pick up five cubes with different alphabet letters printed on them and place them on corresponding docks. The cubes are initially positioned on a shelf within the virtual room. Participants must use their virtual hand or controller (depending on the mode) to grasp the cubes by pressing the trigger button and moving them to the correct dock that matches the letter printed on the cube. The cubes are deliberately placed at varying heights on the shelf to simulate different difficulty levels in picking them up, encouraging participants to extend their arms when necessary. This task assesses various VR interactions involving object manipulation and spatial movement within the virtual scene.

\paragraph{\textbf{Cannon Task}}
Virtual balls are launched into the virtual environment by different cannon shooters which users must catch. Participants are positioned behind a virtual line, and the cannon shooters appear on a wall before them. A total of twenty-four balls are sequentially shot, each with a slightly different direction and velocity. Participants must attempt to catch the balls using their virtual hands or controllers (depending on the mode), without the need for pressing any buttons—simply by touching the balls. A sound accompanies successful catches, while missed catches trigger a distinct, identifiable sound. This task assesses fast-paced mid-air interactions, requiring quick reflexes and hand-eye coordination.

\paragraph{\textbf{Painting Task}}
Participants had to draw simple shapes on a virtual whiteboard. A template of basic figures is displayed on the whiteboard, including a triangle, a square, and a circle. Using a virtual brush or the tip of their virtual index finger (in the absence of controllers), participants approach the whiteboard surface to create strokes and trace the outlines of the given shapes. This task assesses participants' precision in interacting with the virtual environment by following the template and accurately reproducing the shapes through drawing actions.

\section{User Study}
To address our research questions, we conducted a within-subjects study encompassing five conditions corresponding to the five different interaction modes in VR. While the primary independent variable in our study is the interaction modes, we also analyze the three factors—$\mathit{Controller}$, $\mathit{Attached}$, and $\mathit{Stretch}$—as separate independent variables to discover additional insights.

\subsection{Apparatus}
The experiments took place within a laboratory room measuring 6\,m x 6\,m in size. The average duration of the experiment was approximately 45 minutes. The virtual environment was developed using Unity 2021.3 and executed on a PC equipped with an Intel Core i7-12700k CPU, an Nvidia GeForce RTX 3090 GPU, and 32\,GB of RAM. For the VR experience, an HTC Vive Pro HMD with a resolution of 1440\,x\,1600 pixels per eye, a field of view of 110º, and a refresh rate of 90\,Hz was utilized. An external battery and a wireless adapter were employed for the HTC Vive system to enhance freedom of movement, eliminating the need for cables. The participants' pelvis and feet were tracked using three 6-DoF HTC Vive trackers 3.0, while two HTC Vive controllers were held in the participants' hands. In order to minimize occlusions, four SteamVR Base Station 2.0 units were installed at each corner of the room.

\subsection{Participants}

A total of 40 participants took part in the study: 34 were right-handed and 6 left-handed, with a gender distribution of 19 females and 21 males. 
Most participants were university students between the ages of 18 and 24. Participants were not compensated. Regarding gaming experience, 14 participants reported high, 13 medium, 10 low, and 4 no experience. Regarding VR experience, 6 reported high, 3 medium, 18 low, and 14 no experience. Nearly all participants were familiar with using computers, although only a few had prior exposure to VR technology.

The Ethics Committee of the Universitat Politècnica de Catalunya (UPC) issued a favorable opinion on the ethical aspects related to the research carried out in this project (ID Code: 2023.04). The favorable approval of the application implies that the reviewed project complies with the criteria established by the institution's own expertise.

\subsection{Design}
\label{sec:user_study:design}

\begin{figure*}[ht]
  \centering
  \includegraphics[width=1.0\linewidth]{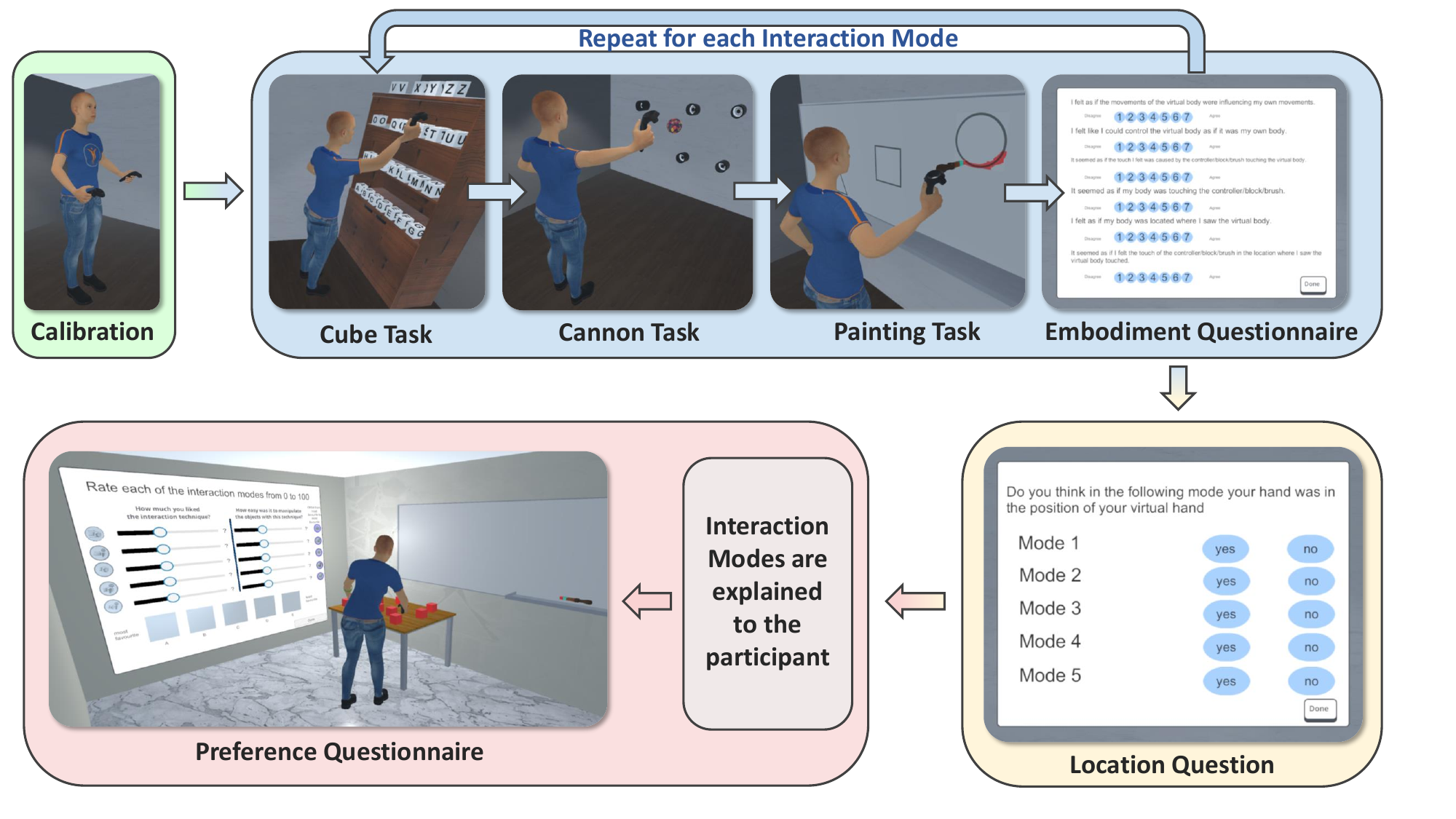}
  \caption{Diagram of the user study protocol. The experiment begins with participants being embodied in a virtual avatar, animated using the \emph{AvatarGo} library \citep{Ponton:2022}. Participants then perform three tasks (in randomized order) in the virtual environment and answer an embodiment questionnaire for each condition. In the subsequent stage, participants answer a question regarding their perception of the positions of their real hands and virtual hands. Finally, after being introduced to the interaction modes, participants explore all conditions freely and provide feedback via the final preference questionnaire.}
  \Description{
  Diagram of the user study protocol. The experiment begins with participants being embodied in a virtual avatar, animated using the AvatarGo library. Participants then perform three tasks (in randomized order) in the virtual environment and answer an embodiment questionnaire for each condition. In the subsequent stage, participants answer a question regarding their perception of the positions of their real hands and virtual hands. Finally, after being introduced to the interaction modes, participants explore all conditions freely and provide feedback via the final preference questionnaire.
  }
  \label{fig:design_protocol}
\end{figure*}

An overview of the user study protocol can be seen in Fig.~\ref{fig:design_protocol}. At the beginning of the study, participants were embodied in a virtual avatar animated using the \emph{AvatarGo} library \citep{Ponton:2022}, which was designed to induce avatar embodiment. Having a well-calibrated avatar already produces high levels of embodiment and is beneficial for interaction in VR \citep{dewez:2021}. \np{AvatarGo employs Unity's built-in IK solver to animate the avatar's limbs (2-segment kinematic chains). However, their method does not solve each limb considering the full-body pose. Instead, it computes the positions of each limb's joints independently, targeting one end-effector per limb. To enhance the overall body pose, AvatarGo integrates forward kinematics (FK) for two additional joints\textemdash the head and spine. This allows the self-avatar to perform more complex movements, such as leaning forwards and sideways, contributing to a more realistic and dynamic representation.} Therefore, participants could see and move through the virtual world as if they were controlling a virtual person with their own movements, and focus on the interaction techniques rather than other embodiment issues. Furthermore, participants were given sufficient time to adjust and become accustomed to their new virtual avatars by following a series of standard embodiment instructions.

Throughout the first part of the experiment, for each condition, participants played in the virtual environment the three different tasks explained in Section~\ref{sec:tasks}: the $\mathit{Cube}$, $\mathit{Cannon}$ and $\mathit{Painting}$ tasks. The order of the conditions was established by a 10\,x\,5 balanced Latin square, and within each condition, the order in which the tasks were performed was randomized. After each condition, an embodiment questionnaire (see Table~\ref{tab:embodimentQuestionnaire}) was displayed inside the virtual environment as shown in Fig.~\ref{fig:design_protocol}. Questions were based on the work by \citet{peck:2021} in which they propose four key embodiment components: appearance, response, ownership, and multi-sensory. We selected a subset of six questions to suit the specific needs of our experiment, with the aim of avoiding participant fatigue considering the duration of the experiment. While the chosen questions cover all the proposed four components of the SoE, they were particularly focused on those highly influencing ownership and multi-sensory aspects.

\begin{table}
    \centering
    \begin{tabular}{ |c|l| } 
         \hline
         \multirow{2}{*}{\textbf{EQ1}} & I felt as if the movements of the virtual body were \\
         & influencing my own movements.  \\ 
         \hline
         \multirow{2}{*}{\textbf{EQ2}} & I felt like I could control the virtual body as if it were \\
         & my own body.  \\ 
         \hline
         \multirow{2}{*}{\textbf{EQ3}} & It seemed as if the touch I felt was caused by the \\
         & controllers/block/brush touching the virtual body.  \\ 
         \hline
         \multirow{2}{*}{\textbf{EQ4}} & It seemed as if my body was touching the \\
         & controller/block/brush.  \\ 
         \hline
         \multirow{2}{*}{\textbf{EQ5}} & I felt as if my body was located where I saw the \\
         & virtual body. \\
         \hline
         \multirow{3}{*}{\textbf{EQ6}} & It seemed as if I felt the touch of the \\
         & controller/block/brush in the location where I saw the \\
         & virtual body touched. \\
         \hline 
    \end{tabular}
    \caption{Embodiment Questionnaire. Participants had to score from 1 to 7 on the statements where 1 means they strongly disagree and 7 means they agree completely.}
    \label{tab:embodimentQuestionnaire}
\vspace*{-10pt}
\end{table}

Following the three tasks across all conditions, we conducted an additional iteration of the conditions. During this phase, participants were queried about their perception of any differences between the positions of their real hands and those of the virtual hands or controllers. To gather this information, a proprioceptive questionnaire was presented within the virtual environment, allowing participants to provide binary responses for each condition to the following question
\textbf{LQ1:} \emph{Do you think your hand was in the position of your virtual hand?}

\begin{figure}[htb]
  \centering
  \includegraphics[width=1.0\linewidth]{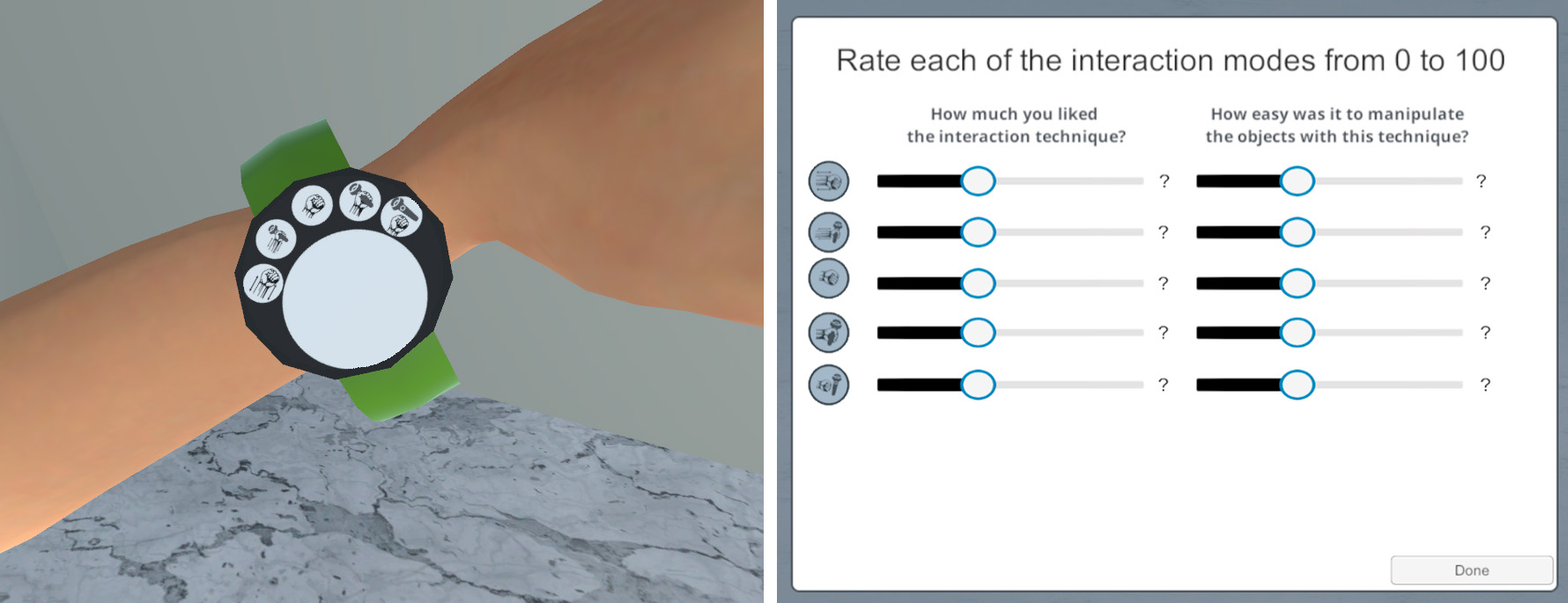}
  \caption{In the preference questionnaire phase, participants could freely switch between interaction methods using the VR controller's touchpad interfaced with a virtual watch display (left), aiding users in visualizing their selected mode. To simplify the recognition of each interaction mode, they were depicted by a unique icon, consistent with its representation in the preference questionnaire (right).}
  \Description{This figure is divided into two parts. On the left, there is an image of a virtual reality (VR) controller's touchpad that interfaces with a virtual watch display. The watch face shows five icons arranged in a circle, each representing different interaction modes. On the right, there is a snapshot of a preference questionnaire. It has two sliders for each interaction mode, asking users to rate from 0 to 100 how much they liked the interaction technique and how easy it was to manipulate the object with the technique. Each slider is accompanied by the corresponding icon from the virtual watch, ensuring consistency between the VR interface and the questionnaire for easier recognition.}
  \label{fig:watch}
\end{figure}

At this stage, the experimenter explained the various interaction techniques to the participants. Following this, participants had the opportunity to freely explore and experience all the conditions while providing their feedback through a final preference questionnaire inside the VR (see Table~\ref{tab:preferenceQuestionnaire} and Fig.~\ref{fig:watch}~right). Notice that participants could change between different modes as many times as they wished using the VR controller's touchpad and a virtual watch (see Fig.~\ref{fig:watch}~left), thus, they could easily compare between interaction modes and rate them based on preference. This questionnaire aimed to capture participants' personal preferences regarding the different interaction techniques used in the study.

\begin{table}
    \centering
    \begin{tabular}{ |c|l| } 
         \hline
         \multirow{2}{*}{\textbf{PQ1}} & How much you liked the interaction technique? \\
         & (Rate each of the interaction modes from 0 to 100) \\
         \hline
         \multirow{3}{*}{\textbf{PQ2}} & How easy was it to manipulate the objects with this \\
         & technique? (Rate each of the interaction modes from \\ 
         & 0 to 100) \\ 
         \hline
    \end{tabular}
    \caption{Preference Questionnaire. Participants had to score from 0 to 100 for each interaction mode for two questions
    .}
    \label{tab:preferenceQuestionnaire}
\vspace*{-10pt}
\end{table}

\subsection{Measures}

In this section, we detail the variables used in the subsequent analysis of the results. First, as suggested by \citet{peck:2021}, we aggregate the embodiment questions to derive the final embodiment score $\mathit{Embodiment} = EQ1 + EQ2 + EQ3 + EQ4 + EQ5 + EQ6$ (see Table~\ref{tab:embodimentQuestionnaire}). Similarly, we calculate user preference as $\mathit{Preference} = PQ1 + PQ2$ (see Table~\ref{tab:preferenceQuestionnaire}). Lastly, we define $\mathit{Proprioception}$ to represent whether users perceived the virtual hand to align with the real one. This binary variable is directly derived from question $LQ1$ (see Section~\ref{sec:user_study:design}).

Following this, we delineate the performance metrics used in each task:
\begin{itemize}
    \item $\mathit{Performance_{Cube}}$ corresponds to the completion time from the start of the \emph{Cube Task} until the final cube is positioned on the table.
    \item $\mathit{Performance_{Cannon}}$ represents the count of balls correctly caught during the \emph{Cannon Task}.
    \item $\mathit{Performance_{Painting}}$ measures the similarity between the overlap user-drawn texture $D$ with the ground truth $G$ in the \emph{Painting Task}. It is computed as the intersection over union of drawn pixels, expressed as $\mathit{Performance_{Painting}} = \frac{G \cap D}{G \cup D}$.
\end{itemize}
Finally, we aggregate the rank transformations of each task's performance metric to compute an overall performance score: 
\begin{align}
\mathit{Performance} = & \frac{1}{3} \big( rank(\mathit{Performance}_{\mathit{Cube}}) \\
& + rank(\mathit{Performance}_{\mathit{Cannon}}) \notag \\
& + rank(\mathit{Performance}_{\mathit{Painting}}) \big) \notag.
\end{align}
Here, for each user, the rank transformation converts numerical values into ranks for each condition, assigning values from 1 (for the least performant condition) to 5 (for the most performant condition).

\subsection{Hypotheses}

In this section, we aim to provide a comprehensive presentation of our hypotheses, outlining the specific aspects of the user study that will be analyzed. Our \emph{a priori} hypotheses are based on previous related work, and are systematically categorized based on the factors under investigation, thereby facilitating the understanding of the results.

\subsubsection{Embodiment} Our first set of hypotheses is based on early work on the effect of visual-sensorimotor contingencies, visual features, and proprioception on the SoE \citep{kilteni:2012b, sanchez-vives:2010, lougiakis:2020, Maselli:2013, botvinick:1998, kilteni2015over}. \textbf{H1} suggests that aligning visual, tactile, and proprioceptive stimuli enhances the SoE. 
Therefore, rendering VR controllers and matching their position (absolute and relative) increases embodiment. Because of that, we hypothesize that not all modes will convey the same level of embodiment (\textbf{H2}). Finally, we also speculate that we can positively affect embodiment if the user believes the controller position is correct (\textbf{H3}).

\begin{itemize}
    \item[] \textbf{H1}\quad $\mathit{Embodiment}$ will be significantly enhanced by \emph{Controller}~(\textbf{H1A}), \emph{Attached}~(\textbf{H1B}), \emph{Stretch}~(\textbf{H1C}) and \emph{Location} (\textbf{H1D}).
    \item[] \textbf{H2}\quad Users will have different degree of $\mathit{Embodiment}$ depending on the interaction mode ($\mathit{Mode}$).
    \item[] \textbf{H3}\quad $\mathit{Embodiment}$ will be positively influenced by the perceived location ($\mathit{Proprioception}$).
\end{itemize}

\subsubsection{Perceived Location} Hypothesis \textbf{H4} suggests that rendering the controller can negatively affect proprioception, whereas having correct multisensory feedback can positively affect it. The possible misalignment between the physical and real location due to the visuo-tactile inconsistency can cause breaks in the embodiment \citep{Burns:2005}.
Conversely, conditions that effectively align visual and tactile stimuli are expected to influence the perceived location accuracy positively \citep{slater:2008, botvinick:1998, sanchez-vives:2010}.

\begin{itemize}
    \item[] \textbf{H4}\quad $\mathit{Proprioception}$ will be negatively affected by \emph{Controller} (\textbf{H4A}), and positively affected by $\mathit{Attached}$~(\textbf{H4B}), $\mathit{Stretch}$ (\textbf{H4C}) and $\mathit{Location}$~(\textbf{H4D}).
\end{itemize}

\subsubsection{Preference} Similarly to the initial set of hypotheses (\textbf{H1-H3}), we propose that the different conditions will enhance user preference (\textbf{H5}). Although the initial set of hypotheses is based on previous work on the SoE, we hypothesize that increasing the SoE will benefit user preference similarly to the work by \citet{fribourg:2020}. Consequently, varying preference levels depend on the interaction mode (\textbf{H6}), and increasing embodiment and perceived location will positively impact user preference (\textbf{H7}). 

\begin{itemize}
    \item[] \textbf{H5}\quad $\mathit{Preference}$ will be significantly enhanced by $\mathit{Controller}$ (\textbf{H5A}), $\mathit{Attached}$~(\textbf{H5B}), $\mathit{Stretch}$~(\textbf{H5C}) and $\mathit{Location}$~(\textbf{H5D}).
    \item[] \textbf{H6}\quad Users will have different degree of $\mathit{Preference}$ depending on the interaction mode ($\mathit{Mode}$).
    \item[] \textbf{H7}\quad $\mathit{Preference}$ will be positively influenced by $\mathit{Embodiment}$ (\textbf{H7A}) and $\mathit{Proprioception}$~(\textbf{H7B}).
\end{itemize}

\subsubsection{Performance} Finally, the last set of hypotheses examines the impact of the different conditions on task performance. Previous studies found that higher performance in some tasks can be achieved when the SoE is high \citep{ries_Analyzingeffectvirtual_2009,pan_AvatarTypeAffects_2019,pan_HowFootTracking_2019}. Therefore, we hypothesize that the same factors that contribute to the SoE will improve performance. We investigate each task individually (\textbf{H8-13}). Lastly, we propose that increased embodiment, perceived location, and preference will positively enhance task performance (\textbf{H14}).

\begin{itemize}
    \item[] \textbf{H8-H10}\quad $\mathit{Performance_{Cube}}$~(\textbf{H8}), $\mathit{Performance_{Cannon}}$~(\textbf{H9}) and $\mathit{Performance_{Painting}}$~(\textbf{H10}) will be significantly enhanced by $\mathit{Controller}$~(\textbf{H8A}-\textbf{H10A}), $\mathit{Attached}$~(\textbf{H8B}-\textbf{H10B}), $\mathit{Stretch}$ (\textbf{H8C}-\textbf{H10C}) and $\mathit{Location}$~(\textbf{H8D}-\textbf{H10D}).
    \item[] \textbf{H11-H13}\quad Users will have different degree of $\mathit{Performance_{Cube}}$ (\textbf{H11}), $\mathit{Performance_{Cannon}}$~(\textbf{H12}) and $\mathit{Performance_{Painting}}$ (\textbf{H13}) depending on the interaction mode ($\mathit{Mode}$).
    \item[] \textbf{H14}\quad $\mathit{Performance}$ will be positively influenced by \emph{Embodiment} (\textbf{H14A}), $\mathit{Proprioception}$~(\textbf{H14B}) and $\mathit{Preference}$~(\textbf{H14C}).
\end{itemize}

\section{Results}

In this section, we present an overview of the results obtained from the statistical analysis and revise whether the presented hypotheses are substantiated or refuted. For an in-depth discussion and interpretation of these findings, please refer to Section~\ref{sec:discussion}.

Shapiro-Wilk tests indicated significant deviations from normality in some instances. As a result, all analyses are carried out using non-parametric tests. To examine the influence of the four conditions\textemdash $\mathit{Controller}$, $\mathit{Attached}$, $\mathit{Stretch}$, $\mathit{Location}$\textemdash on $\mathit{Embodiment}$, $\mathit{Preference}$, and the performance metrics, we employ Wilcoxon tests. Usage of ANOVA was ruled out due to the insufficient combinations available for the study. However, to mitigate the risk of Type 1 errors, we adjust the p-values using the Bonferroni correction. We also report the Wilcoxon effect size ($r$). We did not observe significant differences based on varying levels of VR experience among participants, therefore, we do not report the results separately.

When studying the differences between interaction modes, we perform a one-way repeated measures ANOVA on ranks (Friedman test) followed by post-hoc tests based on the Wilcoxon test. Similarly, p-values are adjusted with the Bonferroni correction. We present Kendall's W effect size for the ANOVA and the Wilcoxon effect size ($r$) for post hoc tests.

Lastly, we leverage linear and binomial mixed-effects models to account for the repeated measures on the same subjects. We employ the binomial model to study $\mathit{Proprioception}$ given its binary nature, and linear models to investigate the relationship between $\mathit{Embodiment}$, $\mathit{Preference}$, $\mathit{Proprioception}$, and $\mathit{Performance}$. We scale numerical data when multiple factors are involved for easy comparison.

\np{To gain a deeper understanding of the $\mathit{Stretch}$ effect on the avatar, we also measured the maximum arm stretch distance for each user when the $\mathit{Stretch}$ condition was active. This distance varies among users, as it is dependent on both the individual's arm length and the virtual avatar's proportions. In our study, we found that the average maximum stretch distance reached 26.3\,cm, with a standard deviation of 4.65\,cm.}

\subsection{Embodiment}
\label{sec:res:embodiment}

Table~\ref{tab:factors_anova_embodiment} presents the results from the Wilcoxon tests on $\mathit{Embodiment}$. Notably, $\mathit{Controller}$ and $Attach$ exhibit no significance, and their effect sizes are near zero. On the contrary, both $\mathit{Stretch}$ and $\mathit{Location}$ yield significant results with moderate effect sizes. Consequently, we dismiss H1A-B and accept H1C-D.

\begin{figure*}[htb]
\begin{minipage}{.55\textwidth}
\centering
\includegraphics[width=\linewidth]{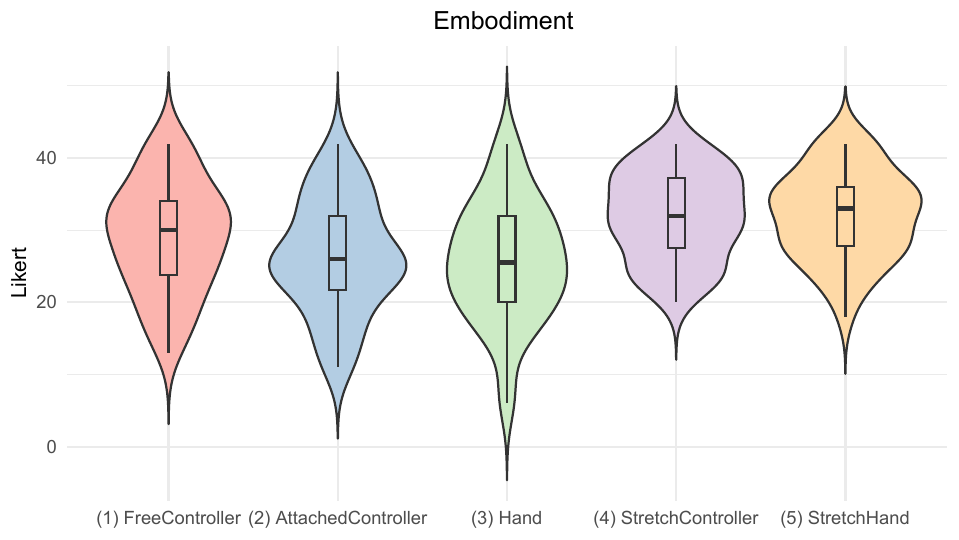}
\end{minipage}%
\begin{minipage}{.4\textwidth}
\centering
\begin{tabular}{lccl}
\hline
Effect & $p$ & Effect Size & \\
\hline
$\mathit{Mode}$& \textbf{<\,.001} & 0.314 & moderate \\
1 - 2 & \textbf{<\,.05}  & 0.454 & moderate \\
1 - 3 & 0.06    & 0.441 & moderate \\
1 - 4 & \textbf{<\,.01}  & 0.554 & large    \\
1 - 5 & \textbf{<\,.05}  & 0.483 & moderate \\
2 - 3 & 1.0     & 0.139 & small    \\
2 - 4 & \textbf{<\,.001} & 0.702 & large    \\
2 - 5 & \textbf{<\,.001} & 0.716 & large    \\
3 - 4 & \textbf{<\,.001} & 0.749 & large    \\
3 - 5 & \textbf{<\,.001} & 0.754 & large    \\
4 - 5 & 1.0     & 0.021 & small    \\
\hline
\end{tabular}
\end{minipage}
\caption{Hypothesis \textbf{H2}. One-way repeated measures ANOVA on ranks (Friedman test) of $\mathit{Mode}$ on $\mathit{Embodiment}$ and the corresponding post-hoc tests (Wilcoxon signed-rank test). $p$ is the adjusted p-value with Bonferroni correction. Effect Size is Kendall's W for the Friedman test and the $r$ value for the post-hoc tests.}
\Description{
Visual representation of the following data:
Effect | p | Effect Size |
-----------------------------------------
Mode | < .001 | 0.314 | moderate
1 - 2 | < .05 | 0.454 | moderate
1 - 3 | 0.06 | 0.441 | moderate
1 - 4 | < .01 | 0.554 | large
1 - 5 | < .05 | 0.483 | moderate
2 - 3 | 1.0 | 0.139 | small
2 - 4 | < .001 | 0.702 | large
2 - 5 | < .001 | 0.716 | large
3 - 4 | < .001 | 0.749 | large
3 - 5 | < .001 | 0.754 | large
4 - 5 | 1.0 | 0.021 | small
----------------------------------------
}
\label{fig:modes_anova_embodiment}
\end{figure*}

Fig.~\ref{fig:modes_anova_embodiment} displays the results of a Friedman test examining the impact of $\mathit{Mode}$ on $\mathit{Embodiment}$. It reveals moderate differences among the groups, implying that certain interaction modes result in elevated $\mathit{Embodiment}$ levels. Hence, we accept H2. In the post-hoc tests, we observe significant effects when contrasting methods that include $\mathit{Stretch}$ with those that do not. However, no distinguishable difference is observed between the two methods that employ $\mathit{Stretch}$.

When we predict $\mathit{Embodiment}$ from $\mathit{Proprioception}$ using a linear mixed-effects model, we notice a positive effect ($\text{Estimate} = 2.275$, $\text{Std. Error} = 0.777$, $t = 2.928$, $\mathbf{p < .01}$). This result aligns with H1C-D, suggesting that when users perceive their hand to be correctly positioned, they experience increased embodiment. Therefore, we accept H3.

\begin{table}[htb]
\begin{tabular}{clccl}
\hline
Hypothesis & Effect & $p$ & $r$ & \\
\hline
H1A & $\mathit{Controller}$ & 0.912 & 0.007 & small \\
H1B & $\mathit{Attached}$  & 0.917 & 0.009 & small \\
\textbf{H1C} & $\mathit{Stretch}$ & \textbf{<\,.001} & 0.311 & moderate \\
\textbf{H1D} & $\mathit{Location}$ & \textbf{<\,.001} & 0.307 & moderate \\
\hline
\end{tabular}
\caption{Wilcoxon tests of $\mathit{Controller}$, $\mathit{Attached}$, $\mathit{Stretch}$ and $\mathit{Location}$ on $\mathit{Embodiment}$. $p$ is the adjusted p-value with Bonferroni correction and $r$ is the Wilcoxon effect size.}
\label{tab:factors_anova_embodiment}
\end{table}

\subsection{Perceived Hand Location (Proprioception)}

\begin{figure}[htb]
  \centering
  \includegraphics[width=0.75\linewidth]{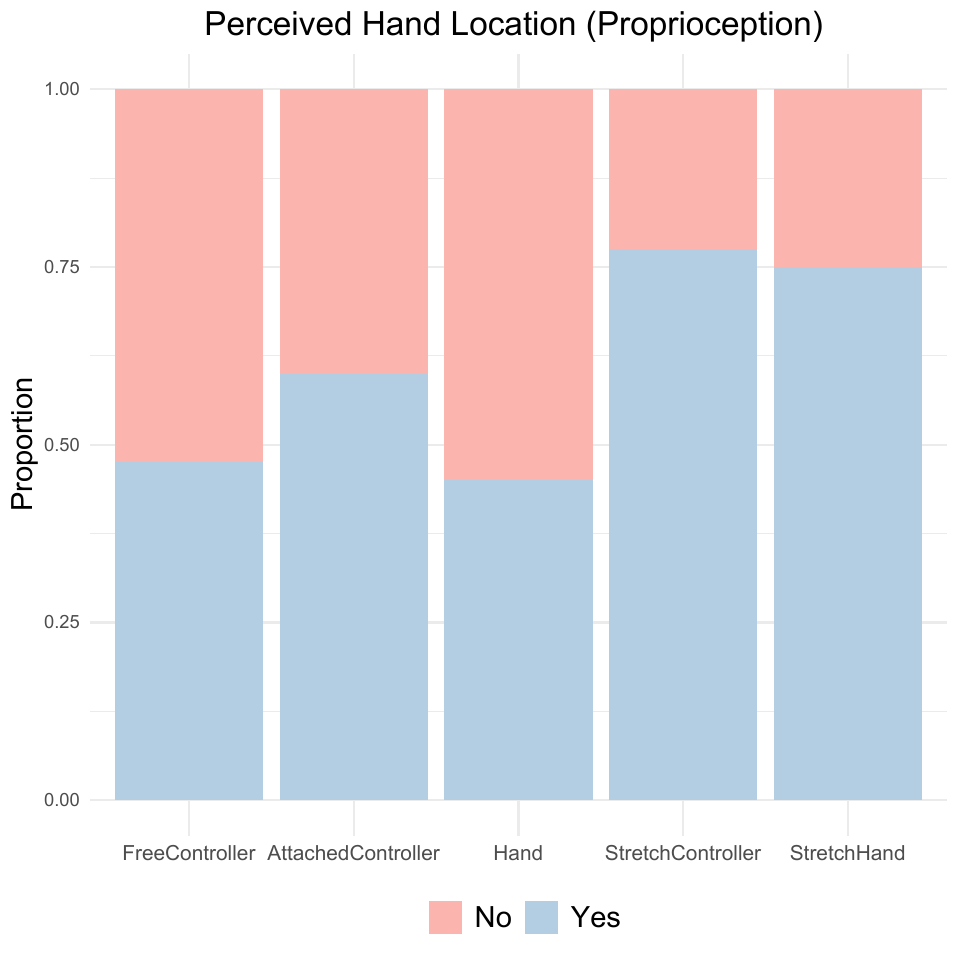}
  \caption{Proportional representation of responses to the question \textbf{LQ1} \emph{Do you think your hand was in the position of your virtual hand?} ($\mathit{Proprioception}$) across different interaction modes ($\mathit{Mode}$). Each mode is represented by a stacked bar, indicating the proportion of \emph{Yes}~(1) and \emph{No}~(0) responses.}
  \Description{Visual representation of the following data:
                FreeController: < 50 percent of people perceived the hand location as correct.
                AttachedController: > 60 percent of people perceived the hand location as correct.
                Hand: < 50 percent of people perceived the hand location as correct.
                StretchController: > 75 percent of people perceived the hand location as correct.
                StretchHand: > 75 percent of people perceived the hand location as correct.}
  \label{fig:perceived_location}
\end{figure}

In Table~\ref{tab:binomial_factors}, the results of the binomial mixed-effects model applied to $\mathit{Proprioception}$ are presented. Notably, $\mathit{Location}$ does not significantly influence the perceived location, as evidenced by its low odds ratio of $1.809$, especially when compared to $\mathit{Attached}$ ($2.430$) and $\mathit{Stretch}$ ($3.104$). It is only $\mathit{Stretch}$ that exhibits a significant impact on $\mathit{Proprioception}$. Therefore, we accept H4C and reject H4A, H4B and H4D. The results can further be examined in Fig.~\ref{fig:perceived_location}, where the proportion of participants answering \emph{Yes} to $LQ1$ exceeded $50\%$ exclusively in the interaction modes incorporating $\mathit{Stretch}$.

\begin{table}[htb]
\begin{tabular}{clccccc}
\hline
Hypo. & Effect & Est. & SE & Odds Ratio & $z$ & $p$ \\
\hline
H4A & $\mathit{Controller}$ & 0.069 & 0.295 & 1.072 & 0.237 & 1.000 \\
H4B & $\mathit{Attached}$  & 0.888 & 0.398 & 2.430 & 2.232 & 0.102 \\
\textbf{H4C} & $\mathit{Stretch}$   & 1.133 & 0.319 & 3.105 & 3.541 & \textbf{<\,.01} \\
H4D & $\mathit{Location}$  & 0.593 & 0.296 & 1.809 & 2.003 & 0.180 \\
\hline
\end{tabular}
\caption{Binomial generalized linear mixed-effects models of $\mathit{Controller}$, $\mathit{Attached}$, $\mathit{Stretch}$ and $\mathit{Location}$ on $\mathit{Proprioception}$. Est. (Estimate) is the coefficient for the predictor in the logistic model. SE is the standard error of the coefficient estimate. Odds Ratio represents the odds of the user perceiving location when the corresponding effect is set to \emph{yes}. $z$ is the z-value used to determine $p$. $p$ is the adjusted p-value with Bonferroni correction.}
\label{tab:binomial_factors}
\end{table}

\subsection{Preference}

\begin{figure*}[htb]
\begin{minipage}{.55\textwidth}
\centering
\includegraphics[width=\linewidth]{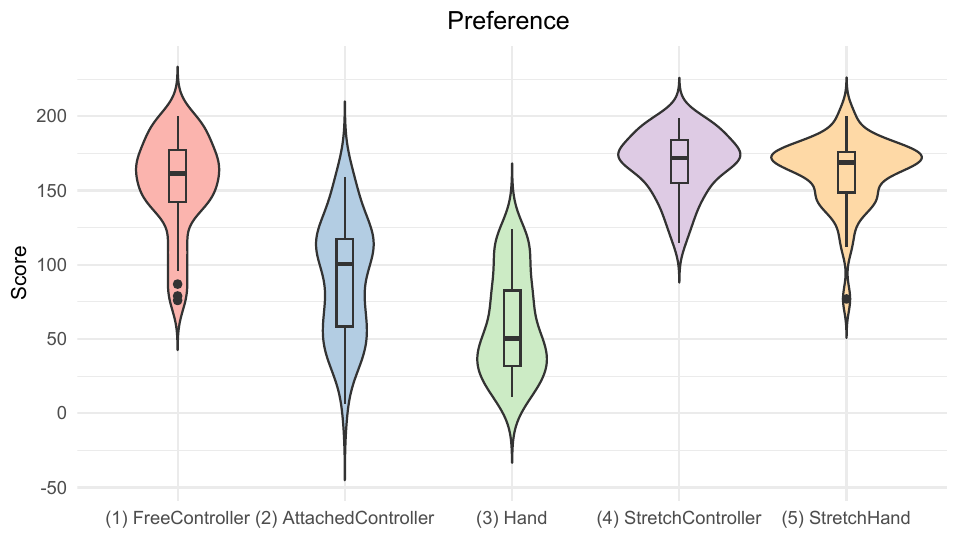}
\end{minipage}%
\begin{minipage}{.4\textwidth}
\centering
\begin{tabular}{lccl}
\hline
Effect & $p$ & Effect Size & \\
\hline
$\mathit{Mode}$& \textbf{<\,.001} & 0.764 & large \\
1 - 2 & \textbf{<\,.001} & 0.860 & large \\
1 - 3 & \textbf{<\,.001} & 0.871 & large \\
1 - 4 & 0.589   & 0.300 & small \\
1 - 5 & 1.0     & 0.174 & small \\
2 - 3 & \textbf{<\,.001} & 0.798 & large \\
2 - 4 & \textbf{<\,.001} & 0.871 & large \\
2 - 5 & \textbf{<\,.001} & 0.850 & large \\
3 - 4 & \textbf{<\,.001} & 0.871 & large \\
3 - 5 & \textbf{<\,.001} & 0.871 & large \\
4 - 5 & 0.997   & 0.267 & small \\
\hline
\end{tabular}
\end{minipage}
\caption{Hypothesis \textbf{H6}. One-way repeated measures ANOVA on ranks (Friedman test) of $\mathit{Mode}$ on $\mathit{Preference}$ and the corresponding post-hoc tests (Wilcoxon signed-rank test). $p$ is the adjusted p-value with Bonferroni correction. Effect Size is Kendall's W for the Friedman test and the $r$ value for the post-hoc tests.}
\Description{
Visual representation of the following data:
-----------------------------------------
Effect | p | Effect Size |
-----------------------------------------
Mode | < .001 | 0.764 | large
1 - 2 | < .001 | 0.860 | large
1 - 3 | < .001 | 0.871 | large
1 - 4 | 0.589 | 0.300 | small
1 - 5 | 1.0 | 0.174 | small
2 - 3 | < .001 | 0.798 | large
2 - 4 | < .001 | 0.871 | large
2 - 5 | < .001 | 0.850 | large
3 - 4 | < .001 | 0.871 | large
3 - 5 | < .001 | 0.871 | large
4 - 5 | 0.997 | 0.267 | small
-----------------------------------------
}
\label{fig:modes_anova_preference}
\end{figure*}

Table~\ref{tab:factors_anova_embodiment} outlines the results from the Wilcoxon tests on $\mathit{Preference}$. All factors significantly influenced $\mathit{Preference}$ except for $\mathit{Attached}$. Factors $\mathit{Stretch}$ ($0.588$) and $\mathit{Location}$ ($0.775$) demonstrated large effect sizes. As depicted in Fig.~\ref{fig:modes_anova_preference}, the interaction modes\textemdash \emph{FreeController}, \emph{StretchController}, and \emph{StretchHand}\textemdash that accurately located the controller were preferred. Fig.~\ref{fig:modes_anova_preference} also reveals significant differences between interaction modes, particularly between groups that precisely located the controller and those that did not. Therefore, we accept hypotheses H5A, H5C, H5D, and H6, while H5B is rejected.

Upon predicting $\mathit{Preference}$ from $\mathit{Embodiment}$ ($\text{Estimate} = 0.459$, $\text{Std. Error} = 0.062$, $t = 7.295$, $\mathbf{p < .001}$) and $\mathit{Proprioception}$ ($\text{Estimate}$ $= 0.105$, $\text{Std. Error} = 0.063$, $t = 1.673$, $p = 0.096$) using a linear mixed-effects model, both factors seem to exert a positive impact. However, $\mathit{Embodiment}$ is the only significant linear predictor of $\mathit{Preference}$. This might be due to $\mathit{Proprioception}$ already explaining $\mathit{Embodiment}$, as suggested by hypothesis H3. Thus, we accept H7A and reject H7B.

\begin{table}[htb]
\begin{tabular}{clccl}
\hline
Hypothesis & Effect & $p$ & $r$ & \\
\hline
\textbf{H5A} & $\mathit{Controller}$ & \textbf{<\,.01}  & 0.215 & small \\
H5B & $\mathit{Attached}$  & 0.076   & 0.214 & small \\
\textbf{H5C} & $\mathit{Stretch}$   & \textbf{<\,.001} & 0.588 & large \\
\textbf{H5D} & $\mathit{Location}$  & \textbf{<\,.001} & 0.775 & large \\
\hline
\end{tabular}
\caption{Wilcoxon tests of $\mathit{Controller}$, $\mathit{Attached}$, $\mathit{Stretch}$ and $\mathit{Location}$ on $\mathit{Preference}$. $p$ is the adjusted p-value with Bonferroni correction and $r$ is the Wilcoxon effect size.}
\label{tab:factors_anova_preference}
\end{table}

\subsection{Performance}

Table~\ref{tab:factors_anova_performance} present the results from Wilcoxon tests on $\mathit{Performance_{Cube}}$, $\mathit{Performance_{Cannon}}$ and $\mathit{Performance_{Painting}}$. In the context of the \emph{Cube Task}, all factors with the exception of $\mathit{Controller}$ significantly impacted performance. The largest effect was seen from $\mathit{Location}$ ($0.564$). For the other tasks, the effects were generally smaller. Notably, $\mathit{Attached}$ and $\mathit{Stretch}$ significantly influenced performance in the \emph{Cannon Task}, while only $\mathit{Attached}$ had a significant impact on the \emph{Painting Task}. As a result, we accept H8B-D, H9B-C, and H10B, and reject H8A, H9A, H9D, H10A, and H10C-D.

\begin{figure*}[htb]
\begin{minipage}{.55\textwidth}
\centering
\includegraphics[width=\linewidth]{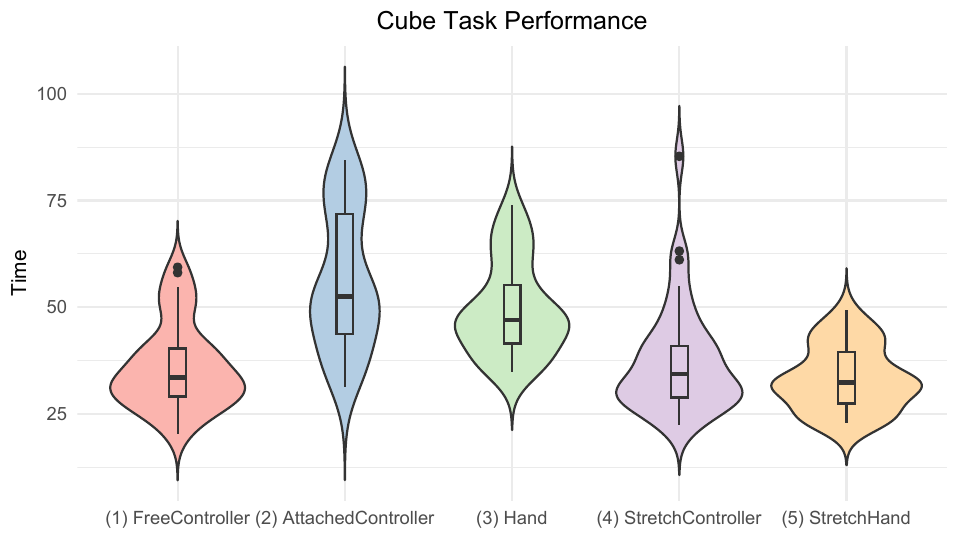}
\end{minipage}%
\begin{minipage}{.4\textwidth}
\centering
\begin{tabular}{lccl}
\hline
Effect & $p$ & Effect Size & \\
\hline
$\mathit{Mode}$& \textbf{<\,.001} & 0.403 & moderate \\
1 - 2 & \textbf{<\,.001} & 0.771 & large \\
1 - 3 & \textbf{<\,.001} & 0.724 & large \\
1 - 4 & 1.0     & 0.167 & small \\
1 - 5 & 1.0     & 0.093 & small \\
2 - 3 & 1.0     & 0.244 & small \\
2 - 4 & \textbf{<\,.01}  & 0.552 & large \\
2 - 5 & \textbf{<\,.001} & 0.667 & large \\
3 - 4 & \textbf{<\,.05}  & 0.482 & moderate \\
3 - 5 & \textbf{<\,.001} & 0.639 & large \\
4 - 5 & 1.0     & 0.199 & small \\
\hline
\end{tabular}
\end{minipage}
\caption{Hypothesis \textbf{H11}. One-way repeated measures ANOVA on ranks (Friedman test) of $\mathit{Mode}$ on $\mathit{Performance_{Cube}}$ and the corresponding post-hoc tests (Wilcoxon signed-rank test). $p$ is the adjusted p-value with Bonferroni correction. Effect Size is Kendall's W for the Friedman test and the $r$ value for the post-hoc tests.}
\Description{
Visual representation of the following data:
-----------------------------------------
Effect | p | Effect Size |
-----------------------------------------
Mode | < .001 | 0.403 | moderate
1 - 2 | < .001 | 0.771 | large
1 - 3 | < .001 | 0.724 | large
1 - 4 | 1.0 | 0.167 | small
1 - 5 | 1.0 | 0.093 | small
2 - 3 | 1.0 | 0.244 | small
2 - 4 | < .01 | 0.552 | large
2 - 5 | < .001 | 0.667 | large
3 - 4 | < .05 | 0.482 | moderate
3 - 5 | < .001 | 0.639 | large
4 - 5 | 1.0 | 0.199 | small
-----------------------------------------
}
\label{fig:modes_anova_performance_cube}
\vspace*{-6pt}
\end{figure*}

\begin{figure*}[htb]
\begin{minipage}{.55\textwidth}
\centering
\includegraphics[width=\linewidth]{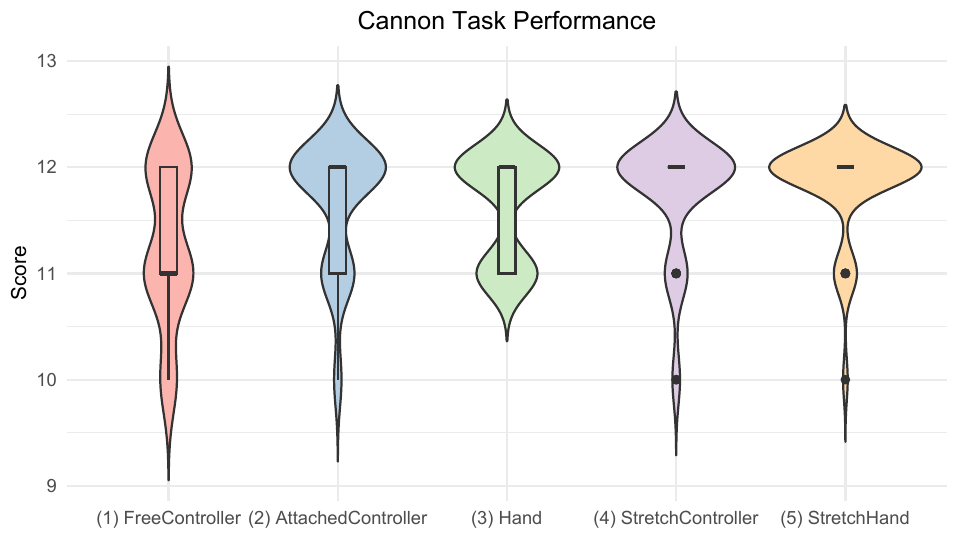}
\end{minipage}%
\begin{minipage}{.4\textwidth}
\centering
\begin{tabular}{lccl}
\hline
Effect & $p$ & Effect Size & \\
\hline
$\mathit{Mode}$& \textbf{<\,.001} & 0.196 & small \\
1 - 2 & 0.117   & 0.402 & moderate \\
1 - 3 & 0.086   & 0.446 & moderate \\
1 - 4 & \textbf{<\,.01}  & 0.629 & large \\
1 - 5 & \textbf{<\,.001} & 0.705 & large \\
2 - 3 & 1.0     & 0.012 & small \\
2 - 4 & 1.0     & 0.207 & small \\
2 - 5 & 0.510   & 0.298 & small \\
3 - 4 & 1.0     & 0.154 & small \\
3 - 5 & 1.0     & 0.283 & small \\
4 - 5 & 1.0     & 0.094 & small \\
\hline
\end{tabular}
\end{minipage}
\caption{Hypothesis \textbf{H12}. One-way repeated measures ANOVA on ranks (Friedman test) of $\mathit{Mode}$ on $\mathit{Performance_{Cannon}}$ and the corresponding post-hoc tests (Wilcoxon signed-rank test). $p$ is the adjusted p-value with Bonferroni correction. Effect Size is Kendall's W for the Friedman test and the $r$ value for the post-hoc tests.}
\Description{
Visual representation of the following data:
-----------------------------------------
Effect | p | Effect Size |
-----------------------------------------
Mode | < .001 | 0.196 | small
1 - 2 | 0.117 | 0.402 | moderate
1 - 3 | 0.086 | 0.446 | moderate
1 - 4 | < .01 | 0.629 | large
1 - 5 | < .001 | 0.705 | large
2 - 3 | 1.0 | 0.012 | small
2 - 4 | 1.0 | 0.207 | small
2 - 5 | 0.510 | 0.298 | small
3 - 4 | 1.0 | 0.154 | small
3 - 5 | 1.0 | 0.283 | small
4 - 5 | 1.0 | 0.094 | small
-----------------------------------------
}
\label{fig:modes_anova_performance_painting}
\vspace*{-10pt}
\end{figure*}

\begin{figure*}[htb]
\begin{minipage}{.55\textwidth}
\centering
\includegraphics[width=\linewidth]{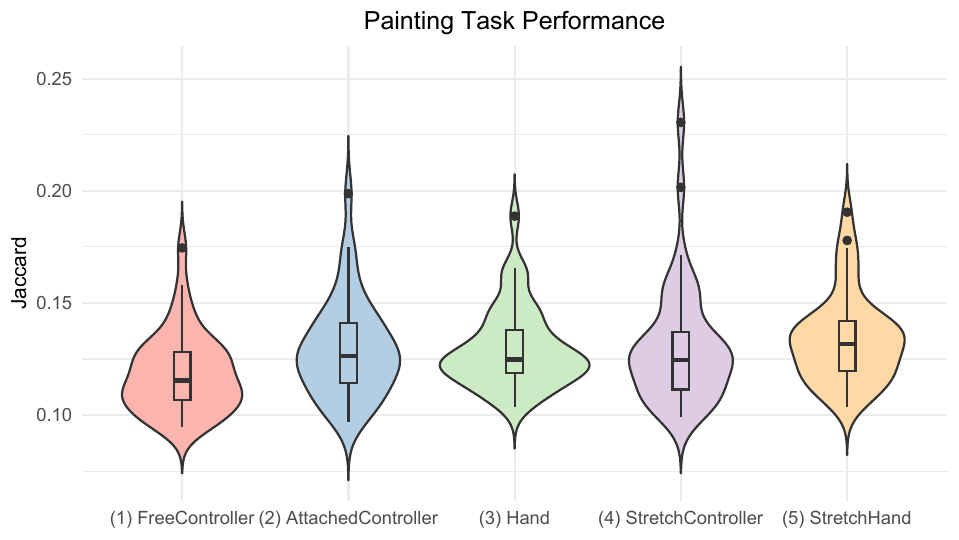}
\end{minipage}%
\begin{minipage}{.4\textwidth}
\centering
\begin{tabular}{lccl}
\hline
Effect & $p$ & Effect Size & \\
\hline
$\mathit{Mode}$& \textbf{<\,.01} & 0.104 & small \\
1 - 2 & \textbf{<\,.05}  & 0.469 & moderate \\
1 - 3 & \textbf{<\,.05}  & 0.448 & moderate \\
1 - 4 & 0.076   & 0.416 & moderate \\
1 - 5 & \textbf{<\,.001} & 0.648 & large \\
2 - 3 & 1.0     & 0.040 & small \\
2 - 4 & 1.0     & 0.034 & small \\
2 - 5 & 1.0     & 0.193 & small \\
3 - 4 & 1.0     & 0.065 & small \\
3 - 5 & 1.0     & 0.212 & small \\
4 - 5 & 1.0     & 0.197 & small \\
\hline
\end{tabular}
\end{minipage}
\caption{Hypothesis \textbf{H13}. One-way repeated measures ANOVA on ranks (Friedman test) of $\mathit{Mode}$ on $\mathit{Performance_{Painting}}$ and the corresponding post-hoc tests (Wilcoxon signed-rank test). $p$ is the adjusted p-value with Bonferroni correction. Effect Size is Kendall's W for the Friedman test and the $r$ value for the post-hoc tests.}
\Description{
Visual representation of the following data:
-----------------------------------------
Effect | p | Effect Size |
-----------------------------------------
Mode | < .01 | 0.104 | small
1 - 2 | < .05 | 0.469 | moderate
1 - 3 | < .05 | 0.448 | moderate
1 - 4 | 0.076 | 0.416 | moderate
1 - 5 | < .001 | 0.648 | large
2 - 3 | 1.0 | 0.040 | small
2 - 4 | 1.0 | 0.034 | small
2 - 5 | 1.0 | 0.193 | small
3 - 4 | 1.0 | 0.065 | small
3 - 5 | 1.0 | 0.212 | small
4 - 5 | 1.0 | 0.197 | small
-----------------------------------------
}
\label{fig:modes_anova_performance_cannon}
\vspace*{-10pt}
\end{figure*}

Figures~\ref{fig:modes_anova_performance_cube},~\ref{fig:modes_anova_performance_cannon}~and~\ref{fig:modes_anova_performance_painting} displays the outcomes of a Friedman test evaluating the influence of $\mathit{Mode}$ on $\mathit{Performance_{Cube}}$, $\mathit{Performance_{Cannon}}$ and $\mathit{Performance_{Painting}}$. Moderate differences among groups were observed for the \emph{Cube Task}, and smaller differences were noted for the other tasks. Despite these differences being primarily small, significant distinctions were identified, leading us to accept hypotheses H11, H12, and H13.

Finally, in the linear mixed-effects model predicting $\mathit{Performance}$ from $\mathit{Embodiment}$ ($\text{Estimate} = 0.086$, $\text{Std. Error} = 0.076$, $t = 1.137$, $p = 0.256$), $\mathit{Proprioception}$ ($\text{Estimate} = 0.139$, $\text{Std. Error} = 0.075$, $t = 1.851$, $p = 0.065$), and $\mathit{Preference}$ ($\text{Estimate} = 0.256$, $\text{Std. Error} = 0.077$, $t = 3.320$, $\mathbf{p < .01}$), all factors appear to contribute positively to $\mathit{Performance}$. However, only $\mathit{Preference}$ significantly predicts $\mathit{Performance}$ in a linear manner, with $\mathit{Proprioception}$ showing a trend toward positive prediction. Consequently, we accept H14C and reject H14A-B.

\begin{table}[htb]
\begin{tabular}{c|clccl}
\hline
Task & Hypothesis & Effect & $p$ & $r$ & \\
\hline
\multirow{4}{*}{Cube} & H8A & $\mathit{Controller}$ & 1.0 & 0.015 & small \\
                      & \textbf{H8B} & $\mathit{Attached}$  & \textbf{<\,.01} & 0.329 & moderate \\
                      & \textbf{H8C} & $\mathit{Stretch}$   & \textbf{<\,.001} & 0.364 & moderate \\
                      & \textbf{H8D} & $\mathit{Location}$  & \textbf{<\,.001} & 0.564 & large \\
\hline
\multirow{4}{*}{Cannon} & H9A & $\mathit{Controller}$ & 0.169 & 0.144 & small \\
                        & \textbf{H9B} & $\mathit{Attached}$  & \textbf{<\,.001} & 0.348 & moderate \\
                        & \textbf{H9C} & $\mathit{Stretch}$   & \textbf{<\,.001} & 0.269 & small \\
                        & H9D & $\mathit{Location}$  & 1.0 & 0.009 & small \\
\hline
\multirow{4}{*}{Painting} & H10A & $\mathit{Controller}$ & 0.051 & 0.176 & small \\
                          & \textbf{H10B} & $\mathit{Attached}$  & \textbf{<\,.05} & 0.247 & small \\
                          & H10C & $\mathit{Stretch}$   & 0.181 & 0.142 & small \\
                          & H10D & $\mathit{Location}$  & 1.0 & 0.065 & small \\
\hline
\end{tabular}
\caption{Wilcoxon tests of $\mathit{Controller}$, $\mathit{Attached}$, $\mathit{Stretch}$ and $\mathit{Location}$ on $\mathit{Performance_{Cube}}$, $\mathit{Performance_{Cannon}}$ and $\mathit{Performance_{Painting}}$. $p$ is the adjusted p-value with Bonferroni correction and $r$ is the Wilcoxon effect size.}
\label{tab:factors_anova_performance}
\end{table}

\section{Discussion and Guidelines}
\label{sec:discussion}

In this section, we aim to understand and interpret the obtained results, while we also try to derive a few guidelines that could help VR designers enhance user embodiment, proprioception, user preference, and/or task performance in their experiences depending on the requirements of their applications. 

\subsection{Embodiment}

\textbf{When the Sense of Embodiment is essential, do not break body continuity.} \textit{Stretch} and \textit{Location} have a statistically significant effect on embodiment.
\textit{Stretch} also affects users' proprioception (perceiving their hand in the correct location).
These results are also consistent with H3, which states that $\mathit{Proprioception}$ positively affects Embodiment. Therefore, we recommend using \textit{Stretch} to guarantee the correct location of end-effectors while providing body continuity, consistent with the results in \citet{dewez:2021}.

\noindent
\textbf{Multisensory coherency with the VR controllers is not necessary to achieve the Sense of Embodiment.}
We expected to observe higher $\mathit{Embodiment}$ when the controller is attached to the hand since it provides multisensory coherence (visual-tactile). However, contrary to our hypothesis, we found a significant moderate effect ($p < .05$, $\text{Effect Size} = 0.454$) on $\mathit{Embodiment}$ being higher with the \emph{FreeController} mode than with \emph{AttachedController}. 
We believe \emph{AttachedController} yielded lower $\mathit{Embodiment}$ due to the inconsistency between visual-tactile and proprioceptive feedback during the \emph{Cube Task}, because users could perceive that their real hands were not located where the virtual hands were being rendered. Note that in Section~\ref{sec:res:embodiment}, we showed that there is a strong correlation between $\mathit{Proprioception}$ and $\mathit{Embodiment}$, which could explain our results. However, \emph{FreeController} had higher embodiment despite the conflicting visual-tactile feedback. Other studies in the literature \citep{steed:2023} also show that multisensory feedback is not always needed to enhance the SoE. 
It is also known that certain body illusions can trick our proprioceptors \citep{kilteni2015over}. Further studies are needed to understand how conflicting sensory feedback could be used to modify own-body representation. This would make it possible to fill the gaps caused by the mismatches in dimension and animation.

\noindent
\textbf{Maximizing the Sense of Embodiment with \emph{StretchController} and \emph{StretchHand}.} Our findings indicate that to enhance the SoE, it is effective to utilize techniques like \emph{StretchController} and \emph{StretchHand}. These methods involve stretching the virtual arms to align with the perceived position of the user's real arms, thereby addressing any mismatches between the user's physical body and their virtual representation while maintaining bodily continuity. Our results also suggest that achieving a high level of SoE does not necessarily require multisensory coherence; thus, stretching the arms is effective across both modes, with (\emph{StretchController}) or without controller rendering (\emph{StretchHand}).

\subsection{Perceived hand location}
\textbf{
Stretch positively influences the correct perception of hand location.}
Regarding the perceived hand location or $\mathit{Proprioception}$, most users respond affirmatively to the location of their hands being correct when $\mathit{Stretch}$ is enabled since the location of the virtual hand is correctly aligned with the real hand at all times, and body continuity is preserved. Therefore, employing \emph{StretchController} or \emph{StretchHand} is recommended to maximize $\mathit{Proprioception}$. In contrast, while \emph{FreeController} accurately aligns the virtual controller with the real-world counterpart, it negatively impacts $\mathit{Proprioception}$, and thus, should be avoided.

\noindent
\textbf{
Attached controllers can induce a proprioceptive drift.}
When $\mathit{Attached}$ is enabled, we increase the chances of people believing that the hand is in the correct position, even though it may not be. Fig.~\ref{fig:perceived_location} supports this result (over $60\,\%$ of participants reported that their hands were perceived in the correct location). The binomial linear model indicated that including \emph{Attached} may substantially increase the probability of users believing their hands were in the right position ($\text{Odds Ratio}=2.43$).
This is known as proprioceptive drift, which is induced by having synchronous visual-tactile feedback \citep{Gonzalez:2020}.
However, our result was not statistically significant. We believe this happens because the \emph{Cube Task} included far-away objects that were hard to reach, whereas, in previous studies, the positions were always within reach.
Thus, we recommend attaching controllers for tasks that do not require interaction with objects that are hard to reach. For instance, a dancing application with minimal interaction with the environment will benefit from the \emph{AttachedController} mode.

\subsection{Preference}

\textbf{For interaction tasks, provide VR controllers or hand locations as accurately as possible.} Both $\mathit{Stretch}$ and $\mathit{Location}$ are significant and have large effect sizes on $\mathit{Preference}$. 
But we cannot conclude whether $\mathit{Stretch}$ alone affects $\mathit{Preference}$ because, in our study, $\mathit{Stretch}$ always implies correct $\mathit{Location}$. However, we can conclude that $\mathit{Location}$ is essential since even in the \emph{FreeController} mode (in which the controller flies away breaking body continuity and multisensory coherency) $\mathit{Preference}$ is high. During the \emph{Cube Task}, participants had to interact with objects that were hard to reach, and some of them commented that they felt frustrated when a position appeared to be physically reachable. Still, their virtual arm could not reach it. This led to participants preferring interaction modes that allowed them to interact correctly. Therefore, from a preference perspective, it appears that \emph{FreeController}, \emph{StretchController}, and \emph{StretchHand} are equally preferred by users.

\noindent
\textbf{Maximizing embodiment positively affects preference.} 
As we accepted H7A, creating an experience with self-avatars that enhance embodiment can also increase user preference. Therefore, studies and work focusing on maximizing embodiment will directly positively affect preference.

\subsection{Performance}

\textbf{For tasks with hard-to-reach objects, stretching arms provides the best performance without affecting embodiment.}
In our study, performance differences appeared mostly during the \emph{Cube Task}. In this task, users needed to carefully reach for objects located far away; thus, they could observe the differences between interaction metaphors. 
Performance is higher when the end-effector is in the correct location, which occurs when we stretch the arm or use the flying controller. Similarly to $\mathit{Preference}$, modes \emph{FreeController}, \emph{StretchController}, and \emph{StretchHand} would equally perform for pick-and-place tasks.

The \emph{Cannon task} requires very rapid movements, in which the user was mainly focused on the next ball, that they could hardly observe differences, but even then, performance was worse for the \emph{FreeController} mode. 

For the \emph{Painting Task}, the range of movements was mostly limited to a natural arm position, which did not often trigger the arm stretching or the flying controller. However, the few times these differences were observed led to users performing worse with the \emph{FreeController} mode.  

Therefore, for tasks that require accurate interaction with virtual objects, we recommend virtually stretching the arm to allow for correct end-effector positioning while respecting visual-tactile and proprioceptive coherence. For applications that require manipulation of objects within easy reach or that the interaction requires rapid movements without careful manipulation, the stretching technique would also be valid but not necessary. So, in these cases, having attached controllers would suffice.

\section{Limitations}

This section outlines the primary limitations identified in our study. Firstly, the results are derived from a specific set of tasks, primarily centered around interaction tasks designed to mimic everyday activities in VR. This focus means our findings might not fully represent the variety of experiences in VR applications. For instance, in scenarios without active interaction, users might be less likely to notice the incorrect alignment of virtual and real hands, as suggested by \citet{Ponton:2023}.

The generalizability of our results is also potentially limited by the demographics of our participant sample. Most participants were university students with limited VR experience. While embodiment questionnaires are a widely accepted tool, some participants might have faced challenges understanding and accurately responding to them. We minimized confounding factors, yet certain technical aspects, such as pose animation and the designs of avatars and controllers, could have influenced the outcomes. Additionally, we believe that the skinning of the avatar when stretching the arms can significantly affect the results.

\np{Finally, our study did not investigate every possible combination of factors like $\mathit{Stretch}$ and $\mathit{Location}$ or the case where the virtual arm could be longer than the real arm}. To effectively separate the effects of $\mathit{Stretch}$ and $\mathit{Location}$, we would have needed to include an interaction mode where the arm is stretched but not precisely located in the correct position. For instance, \citet{kilteni:2012b} studies the SoE based on different arm lengths. However, such a mode does not align well with practical VR scenarios, leading us to exclude it from our study. This decision, while rational from a practical standpoint, does limit the comprehensiveness of our findings regarding these factors. \np{Note that if the virtual arm were longer than the real hand, the result would be that the IK would bend at the elbow to reach the end-effector. This situation does not prevent the user from reaching objects, it simply introduces a mismatch in the pose of the arm \cite{Ponton:2023}, but not in the position of the end-effector. Our $\mathit{Stretch}$ was limited to the values needed to fill the mismatch between the virtual and real hand, we have not investigated the tolerable mismatch threshold and its impact on embodiment.}

\section{Conclusions and Future Work}

In this paper, we have studied 5 interaction metaphors for animated full-body avatars. We have focused our efforts on the simulation and rendering of arms, hands, and controllers. Our results suggest that selecting the best interaction metaphor strongly depends on the type of task that the user needs to perform. However, results on all embodiment, performance, preference, and proprioception suggest that the best interaction metaphor is to have stretch arms that can allow the end-effectors to be in the correct position while respecting body continuity. Rendering the controllers does not appear to be relevant, although attaching the controllers to the hand may induce a proprioceptive drift due to the consistent visual-tactile feedback. 

In future work, we would like to study interaction metaphors when performing collaborative tasks, for example in cases where two participants need to pass a virtual object or carry something together. We would also like to experiment with how interaction metaphors could be used for the inclusiveness of people with mobility limitations, by improving their interaction in VR beyond their possibilities in the real world. For example, extending the stretch interaction mode to help them reach for objects that are slightly beyond their physical reach, without breaking embodiment or body continuity.

\begin{acks}
This work has received funding from the European Union’s Horizon 2020 research and innovation programme under HORIZON-CL4-2022-HUMAN-01 grant agreement No 101093159 (XR4ED), and from MCIN/AEI/10.13039/501100011033/FEDER "A way to make Europe", UE (PID2021-122136OB-C21). Jose Luis Ponton was also funded by the Spanish Ministry of Universities (FPU21/01927).
\end{acks}


\bibliographystyle{ACM-Reference-Format}
\bibliography{References.bib}



\end{document}